\documentclass{appolb}
\usepackage{graphicx}
\usepackage{times}
\usepackage[english]{babel}
\usepackage{amsmath}
\usepackage[latin1]{inputenc}
\usepackage[T1]{fontenc}
\usepackage{graphicx}
\usepackage{epsfig}
\usepackage{rotating}
\usepackage{subfigure}
\usepackage{tikz}
\usepackage[subfigure]{ccaption}
\usepackage{mathtools,cancel}

\def\ii{\'{\i}}

\begin{document}
\title{Role of current quark mass dependent multi-quark interactions in low lying meson mass spectra%
\thanks{Presented at the workshop "Excited QCD 2013",
03-09 February 2013 in Bjelasnica Mountain, Sarajevo, Bosnia. Work supported by Centro de F\ii sica Computacional, FCT, CERN/FP/116334/2010, 
QREN, UE/FEDER through COMPETE. 
Part of the EU Research Infrastructure Integrating Activity 
Study of Strongly Interacting Matter (HadronPhysics3) under the 7th Framework 
Programme of EU: Grant Agreement No. 283286.}%
}
\author{A. A. Osipov, B. Hiller\footnote{Speaker}, A. H. Blin
\address{Departamento de F\ii sica, CFC, Faculdade de Ci\^encias e Tecnologia da Universidade de Coimbra, P-3004-516 Coimbra, Portugal}
}
\maketitle
\begin{abstract}
We call attention to a class of current-quark mass dependent multi-quark interaction terms which break explicitly the chiral $SU(3)_L\times SU(3)_R$ and $U_A(1)$ symmetries. They complete the set of effective quark interactions that contribute at the same order in $N_c$ as the 't Hooft flavor determinant interaction and the eight quark interactions in the phase of spontaneously broken chiral symmetry. The $N_c$ classification scheme matches the counting rules based on arguments set by the scale of spontaneous chiral symmetry breaking. Together with the leading in $N_c$ four quark Nambu-Jona-Lasinio Lagrangian and current quark mass matrix, the model is apt to account for the correct empirical ordering and magnitude of the splitting of states in the low lying mass spectra of spin zero mesons. The new terms turn out to be essential for the ordering $m_K < m_\eta$ in the pseudoscalar sector and $m_{\kappa_0} < m_{a_0}\sim m_{f0}$ for the scalars.  
\end{abstract}
 
\PACS{PACS: 11.30.Rd; 11.30.Qc; 12.39.Fe; 12.40.Yx; 14.40.Aq; 14.65.Bt}
  
\vspace{0.5cm}

The seminal papers of Nambu and Jona-Lasinio (NJL) \cite{Nambu:1961} more than 50 years ago mark the advent of spontaneous chiral symmetry breaking in particle physics. The recognition of its importance for QCD was established in \cite{Eguchi:1976}. Explicit breaking of the chiral symmetry through a mass term was introduced in \cite{Ebert:1982}. Since then an overwhelming and rich insight has been achieved in the description of non-perturbative effects of QCD through various extensions of the model. One aspect of the model which only recently has been tackled is the role played by the explicit symmetry breaking interacting dynamics \cite{Osipov:2013}. There we argue that quark mass dependent interactions must be introduced at the same level of $N_c$ counting as the $U(1)_A$ breaking 't Hooft determinant \cite{Hooft:1976} extension of the model \cite{Bernard:1988}-\cite{Naito:2003}, and inclusion of eight-quark interactions \cite{Osipov:2006b}, \cite{Osipov:2007a}. 

The scale of applicability of the effective quark Lagrangian is given by the scale of spontaneous breaking of chiral symmetry, of the order of $\Lambda_{\chi SB}\sim 4\pi f_{\pi}$ \cite{Georgi:1984}. In the NJL model this scale is related to the gap equation and given by the ultra-violet cutoff $\Lambda$ of the one-loop quark integral, above which one expects non-perturbative effects to be of less importance. To construct the effective Lagrangian we consider only the generic vertices $L_i$ of non-derivative type and whose contributions to the effective potential survive at $\Lambda\to\infty$      
\begin{equation}
\label{genL}
   L_i\sim \frac{\bar g_i}{\Lambda^\gamma}\chi^\alpha\Sigma^\beta,
\end{equation} 
where powers of $\Lambda$ give the correct dimensionality of the interactions; the $L_i$ are C, P, T and chiral $SU(3)_L\times SU(3)_R$ invariant blocks, built of powers of the  sources $\chi$
which at the end give origin to the explicit symmetry breaking, and of the $U(3)$ Lie-algebra valued field $\Sigma =(s_a-ip_a)
\frac{1}{2}\lambda_a$; here $s_a=\bar q\lambda_aq$, $p_a=\bar q\lambda_ai
\gamma_5q$, and $a=0,1,\ldots ,8$, $\lambda_0=\sqrt{2/3}\times 1$, $\lambda_a$ 
being the standard $SU(3)$ Gell-Mann matrices for $1\leq a \leq 8$. Under 
chiral transformations the quark fields transform as $q'=V_Rq_R+V_Lq_L$, where $q_R=P_R q, q_L=P_Lq$, and 
$P_{R,L}=\frac{1}{2}(1\pm\gamma_5)$. Thus $\Sigma'=V_R\Sigma V_L^\dagger$, and 
$\Sigma^{\dagger'}=V_L\Sigma^\dagger V_R^\dagger$. The transformation property of 
the source is $\chi' =V_R\chi V_L^\dagger$. Finally $\bar g_i=g_i \Lambda^\gamma$ are dimensionless coupling constants. 

One concludes (see \cite{Osipov:2013} for details) that there are (i) only four classes of 
vertices which contribute at $\alpha=0$; those are four, six and eight-quark 
interactions, corresponding to $\beta=2,3$ and $4$ respectively; the $\beta=1$ 
class is forbidden by chiral symmetry requirements; (ii) there are only six 
classes of vertices depending on external sources $\chi$, they are: $\alpha 
=1, \beta =1,2,3$; $\alpha =2, \beta =1,2$; and $\alpha =3, \beta =1$. 

The explicit Lagrangian corresponding to the case (i) is well known
\begin{eqnarray}
\label{L-int}
   L_{int}&=&\frac{\bar G}{\Lambda^2}\mbox{tr}\left(\Sigma^\dagger\Sigma\right)
   +\frac{\bar\kappa}{\Lambda^5}\left(\det\Sigma+\det\Sigma^\dagger\right) 
   \nonumber \\
   &+&\frac{\bar g_1}{\Lambda^8}\left(\mbox{tr}\,\Sigma^\dagger\Sigma\right)^2
   +\frac{\bar g_2}{\Lambda^8}\mbox{tr}
   \left(\Sigma^\dagger\Sigma\Sigma^\dagger\Sigma\right).
\end{eqnarray}  
It contains four dimensionful couplings $G, \kappa, g_1, g_2$.

The second group (ii) contains eleven terms
\begin{equation}
   L_\chi =\sum_{i=0}^{10}L_i,
\end{equation}

\begin{eqnarray}
\label{L-chi-1}
   L_0&=&-\mbox{tr}\left(\Sigma^\dagger\chi +\chi^\dagger\Sigma\right), \hspace{0.5cm}
   L_1=-\frac{\bar\kappa_1}{\Lambda}e_{ijk}e_{mnl}
   \Sigma_{im}\chi_{jn}\chi_{kl}+h.c.
   \nonumber \\
   L_2&=&\frac{\bar\kappa_2}{\Lambda^3}e_{ijk}e_{mnl}
   \chi_{im}\Sigma_{jn}\Sigma_{kl}+h.c., \hspace{0.5cm}
   L_3=\frac{\bar g_3}{\Lambda^6}\mbox{tr}
   \left(\Sigma^\dagger\Sigma\Sigma^\dagger\chi\right)+h.c.
   \nonumber \\
   L_4&=&\frac{\bar g_4}{\Lambda^6}\mbox{tr}\left(\Sigma^\dagger\Sigma\right)
   \mbox{tr}\left(\Sigma^\dagger\chi\right)+h.c., \hspace{0.5cm}
   L_5=\frac{\bar g_5}{\Lambda^4}\mbox{tr}\left(\Sigma^\dagger\chi
   \Sigma^\dagger\chi\right)+h.c.
   \nonumber \\
   L_6&=&\frac{\bar g_6}{\Lambda^4}\mbox{tr}\left(\Sigma\Sigma^\dagger\chi
   \chi^\dagger +\Sigma^\dagger\Sigma\chi^\dagger\chi\right), \hspace{0.5cm}
   L_7=\frac{\bar g_7}{\Lambda^4}\left(\mbox{tr}\Sigma^\dagger\chi 
   + h.c.\right)^2
   \nonumber \\ 
   L_8&=&\frac{\bar g_8}{\Lambda^4}\left(\mbox{tr}\Sigma^\dagger\chi 
   - h.c.\right)^2, \hspace{0.5cm}
   L_9=-\frac{\bar g_9}{\Lambda^2}\mbox{tr}\left(\Sigma^\dagger\chi
   \chi^\dagger\chi\right)+h.c.
   \nonumber \\
   L_{10}&=&-\frac{\bar g_{10}}{\Lambda^2}\mbox{tr}\left(\chi^\dagger\chi\right)
   \mbox{tr}\left(\chi^\dagger\Sigma\right)+h.c.
\end{eqnarray}
The $N_c$ assignments are \cite{Osipov:2013}:
$\Sigma \sim N_c$; $\Lambda \sim N_c^0 \sim 1$; $\chi \sim N_c^0 \sim 1$ \footnote{Note that the counting for $\Lambda$ is a direct consequence of the gap equation $1\sim N_c G\Lambda^2$.}.
Then we get that exactly the diagrams which survive as $\Lambda\rightarrow\infty$ also surive as $N_c\rightarrow\infty$ and comply with the usual requirements, i.e. (i) the leading quark contribution to the vacuum energy from $4q$ interactions is of order $N_c \rightarrow G\sim \frac{1}{N_c}$; 
(ii) the $U_A(1)$ anomaly contribution is suppressed by one power of $\frac{1}{N_c} \rightarrow \kappa \sim \frac{1}{N_c^3}$; 
(iii) Zweig's rule violating effects are always of order $\frac{1}{N_c}$ with respect to the leading contribution.

As a result we have $L_{4q}$ and $L_0$ of ${\cal O}(N_c)$ and all other terms in the Lagrangian are ${\cal O}(N_c^0)$.  
Non OZI-violating Lagrangian pieces scaling as ${\cal O}(N_c^0)$ represent NLO contributions with one internal quark loop in $N_c$ counting. The coupling encodes the admixture of four quark component ${\bar q}q{\bar q}q$ to the leading ${\bar q}q$ at $N_c\rightarrow\infty$.
Diagrams tracing Zweig's rule violation are: $\kappa,\kappa_1,\kappa_2,g_1,g_4,g_7,g_8,g_{10}$;
Diagrams with admixture of 4 quark and 2 quark states are: $g_2,g_3,g_5,g_6,g_9$.

The total Lagrangian is the sum    
\begin{equation}
\label{LQ}
   L=\bar qi\gamma^\mu\partial_\mu q+L_{int}+L_\chi. 
\end{equation}

Finally, having all the building blocks in conformity with the symmetry content of 
the model, one is free to choose the external source $\chi$. Putting $\chi 
={\cal M}/2$, where ${\cal M}=\mbox{diag}(\mu_u, \mu_d, \mu_s)$, 
we obtain a consistent set of explicitly breaking chiral symmetry terms. 
Figure 1 shows a schematic representation of interactions with source $\chi\sim \mu$ insertions. 
\begin{figure}[htb]

\includegraphics[height= 0.5 \textheight,width=0.6 \textwidth,angle=270]{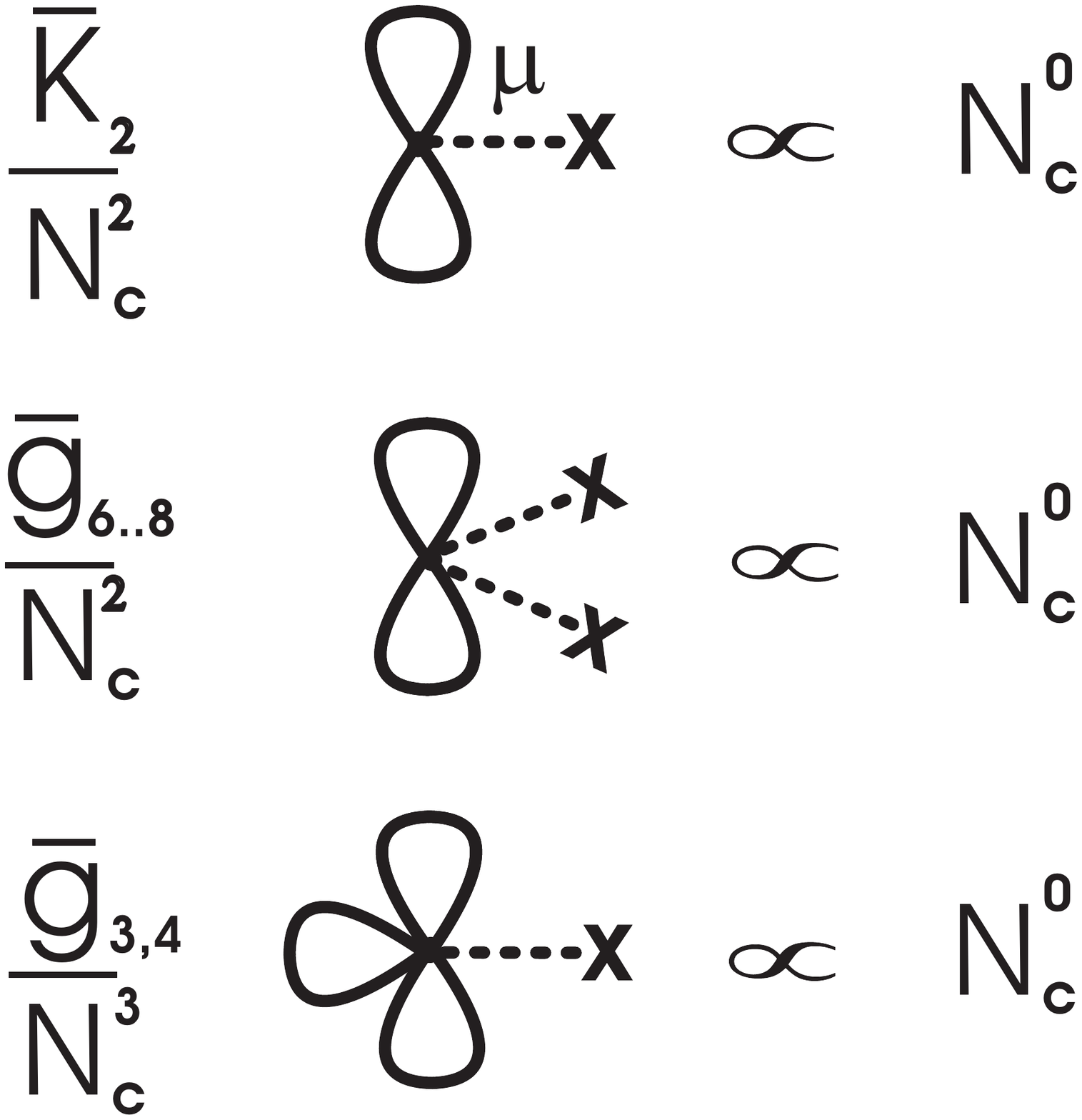}
\caption{\footnotesize Contributions to the effective potential from explicit symmetry breaking interactions, where dashed lines linked to crosses denote the external sources. The $N_c$ dependence carried by the couplings (left) combines with the $N_c$ attributed to each closed quark loop to yield the same counting $N_c^0$ for all interactions. Similar arguments lead to scaling $\sim \Lambda$ for the upper interaction and $\sim \Lambda^0$ for all other.   
}   
\end{figure}

The details of bosonization and regularization which lead finally from (\ref{LQ}) to the long distance mesonic Lagrangian can be found in \cite{Osipov:2013},\cite{Osipov:2001}-\cite{Osipov:1985}.
The model contains eighteen parameters: the scale $\Lambda$, three parameters which
are responsible for explicit chiral symmetry breaking $\mu_u,\mu_d,\mu_s$, 
and fourteen interaction couplings $\bar G, \bar\kappa, \bar\kappa_1, \bar
\kappa_2$, $\bar g_1,\ldots,\bar g_{10}$. Three of them, $\bar\kappa_1, 
\bar g_9, \bar g_{10}$, contribute to the current quark masses and express the Kaplan-Manohar ambiguity \cite{Manohar:1986} within the model. We chose $\kappa_1=g_9=g_{10}=0$ and take the PDG \cite{PDG} current quark 
mass values around $\hat m=4$\ MeV and $m_s=100$\ MeV  to be an input. Seven more 
describe the strength of multi-quark interactions with explicit symmetry 
breaking effects. These vertices contain new details of the quark dynamics 
which had not been studied yet in any NJL-type models.  
The parameter values used to fit the low lying characteristica of the spin $0$ mesons as indicated below in the text are shown in Tables 1 and 2, with two sets which differ by (a) $m_\sigma=600$ MeV, (b) $m_\sigma=500$ MeV. Table 1 contains the usual parameters, which are not much affected by the introduction of the explicit symmetry breaking interactions terms with couplings shown in Table 2. They allow to fit the  pseudoscalar spectrum, $m_\pi=138$\ MeV, 
$m_K=494$\ MeV, $m_\eta=547$\ MeV, $m_{\eta'}=958$\ MeV, the weak pion and kaon 
decay constants, $f_\pi=92$\ MeV, $f_K=113$\ MeV, and the singlet-octet mixing 
angle $\theta_p=-15^\circ$, to perfect accuracy. The calculated 
values of quark condensates are the same for both sets: $-\langle\bar uu
\rangle^{\frac{1}{3}}=232$\ MeV, and $-\langle\bar ss\rangle^{\frac{1}{3}}=206$\ 
MeV.  We stress that without the new explicit symmetry breaking terms this had not been possible. We find that the couplings $g_8$ and $\kappa_2$ are crucial for the high precision within the pseudoscalar sector.      
We find also that the low lying scalar nonet encompassing the $\sigma$ or 
$f_0(600),$ $a_0(980),$ $\kappa (850), f_0(980)$ mesons can be obtained according to the empirical ordering: $m_\kappa<m_{a_0}\simeq m_{f_0}$, with the singlet-octet mixing angle 
$\theta_s$ roughly $\theta_s=19^\circ$. This is in contrast to the $m_\sigma<m_{a_0}<m_\kappa <m_{f_0}$ sequence obtained in the framework of NJL models, e.g. \cite{Weise:1990,Weise:1991,Volkov:1984,Volkov:1986,Osipov:2004b,Osipov:2007a,Su:2007}, see also \cite{Parganlija:2013} in the framework of the linear sigma model.  
The main parameter responsible for a lower mass of $\kappa(800)$ as compared to the mass of $a_0(980)$ is $g_3$; $g_6$ allows for fine tuning. We understand the empirical mass assignment inside the light scalar 
nonet as a consequence of the quark-mass dependent interactions, i.e. as the 
result of some predominance of the explicit chiral symmetry breaking terms over
the dynamical chiral symmetry breaking ones for these states. Note that the couplings $g_3$ and $g_6$ encode 4q and 2q admixtures. This establishes a link between the asymptotic meson states obtained from the effective multiquark interactions considered to the successful approaches which support $\bar qq$ states 
with a meson-meson admixture \cite{Beveren:1986} or mixing of $q\bar q$-states with others, consisting of two quarks and two 
antiquarks, $q^2\bar q^2$  \cite{Jaffe:1977}-\cite{Close:2002}.
  
\begin{table*}
\caption{\small Parameter sets of the model, (a) for $m_\sigma=600$ MeV, (b) for $m_\sigma=500 MeV$: $\hat m, m_s$, and $\Lambda$ are given in 
         MeV. The couplings have the following units: $[G]=$ GeV$^{-2}$, 
         $[\kappa ]=$ GeV$^{-5}$, $[g_1]=[g_2]=$ GeV$^{-8}$. We also show here 
         the values of constituent quark masses $\hat M$ and $M_s$ in MeV.}
\label{table-1}
\begin{tabular*}{\textwidth}{@{\extracolsep{\fill}}lrrrrrrrrl@{}}
\hline
Sets & \multicolumn{1}{c}{$\hat m$} 
     & \multicolumn{1}{c}{$m_s$}
     & \multicolumn{1}{c}{$\hat M$}                         
     & \multicolumn{1}{c}{$M_s$}                             
     & \multicolumn{1}{c}{$\Lambda$}    
     & \multicolumn{1}{c}{$G$}  
     & \multicolumn{1}{c}{$-\kappa$} 
     & \multicolumn{1}{c}{$g_1$}    
     & \multicolumn{1}{c}{$g_2$} \\ 
\hline
a  & 4.0* & 100* & 361 & 526 & 837  & 8.96  & 93.0   & 1534 &0* \\  
b  & 4.0* & 100* & 361 & 526 & 837  & 7.06  & 93.3   & 3420 &0* \\ 
\hline
\end{tabular*} 
\end{table*}   

\begin{table*}
\caption{\small Explicit symmetry breaking interaction couplings. The couplings have
the following units: $[\kappa_1]=$ GeV$^{-1}$, $[\kappa_2]=$ GeV$^{-3}$, 
$[g_3]=[g_4]=$ GeV$^{-6}$, $[g_5]=[g_6]=[g_7]=[g_8]=$ GeV$^{-4}$,
$[g_9]=[g_{10}]=$ GeV$^{-2}$.}
\label{table-2}
\begin{tabular*}{\textwidth}{@{\extracolsep{\fill}}lrrrrrrrrrl@{}}
\hline
Sets  & \multicolumn{1}{c}{$\kappa_1$}
      & \multicolumn{1}{c}{$\kappa_2$}    
      & \multicolumn{1}{c}{$-g_3$}  
      & \multicolumn{1}{c}{$-g_4$} 
      & \multicolumn{1}{c}{$g_5$} 
      & \multicolumn{1}{c}{$-g_6$}   
      & \multicolumn{1}{c}{$-g_7$} 
      & \multicolumn{1}{c}{$g_8$} 
      & \multicolumn{1}{c}{$g_9$}
      & \multicolumn{1}{c}{$g_{10}$} \\ 
\hline
a  &0* & 9.05  & 4967 & 661 & 192.2 & 1236  & 293  & 52.2  &0* &0*  \\  
b  &0* & 9.01  & 4990 & 653 & 192.5 & 1242  & 293  & 51.3  &0* &0*  \\ 
\hline
\end{tabular*}
\end{table*}   
Our result for the mass spectrum of the scalar sector, being promising by itself, 
must however be supported by further processes, such as strong and radiative decays. 
Work in this direction is in progress. Preliminary results point to strong interactions widths 
$\Gamma_{\sigma\pi\pi}\sim 600$ MeV, $\Gamma_{f_0\pi\pi}\sim 60$ MeV,  as well as the two photon decay  $\Gamma_{f_0\gamma\gamma}\sim 0.3$ KeV being within the empirical range \cite{PDG}, while 
$\Gamma_{\sigma\gamma\gamma}\sim 1.2$ KeV is slightly below.


\begin{thebibliography}{99}

\bibitem{Nambu:1961}Y. Nambu, Phys. Rev. 
Lett. {\bf 4}, 380 (1960);
 Y. Nambu, G. Jona-Lasinio, Phys. Rev. {\bf 122}, 
    345 (1961); Y. Nambu, G. Jona-Lasinio, {\bf 124}, 246 (1961); 
\bibitem{Eguchi:1976} T. Eguchi, Phys. Rev. D {\bf 14}, 2755 (1976); K.Kikkawa, Prog. Theor. Phys. {\bf 56}, 947 (1976). 
\bibitem{Ebert:1982} M.K. Volkov, D. Ebert, Sov. J. Nucl. Phys. {\bf 36}, 1265 (1982).
\bibitem{Osipov:2013} A. A. Osipov, B. Hiller, A. H. Blin, Europ. Phys. J. A {\bf 49}, 14 (2013).
\bibitem{Hooft:1976} G. 't Hooft, Phys. Rev. D {\bf 14}, 3432 (1976);
    G. 't Hooft, Phys. Rev. D {\bf 18}, 2199 (1978).
\bibitem{Bernard:1988} V. Bernard, R. L. Jaf\mbox{}fe and
    U.-G. Meissner, Phys. Lett. B {\bf 198}, 92 (1987);   
    V. Bernard, R. L. Jaf\mbox{}fe and U.-G. Meissner, Nucl. Phys. B
    {\bf 308}, 753 (1988). 
\bibitem{Reinhardt:1988} H. Reinhardt and R. Alkofer, Phys. Lett. B 
    {\bf 207}, 482 (1988).
\bibitem{Weise:1990} S. Klimt, M. Lutz, U. Vogl, W. Weise,
   Nucl. Phys. A \textbf{516}, 429 (1990).
\bibitem{Weise:1991} U. Vogl, W. Weise, Progr. Part. Nucl. Phys. \textbf{27}, 
   195 (1991). 
\bibitem{Takizawa:1990} M. Takizawa, K. Tsushima, Y. Kohyama, K. Kubodera, 
   Nucl. Phys. A \textbf{507}, 611 (1990).
\bibitem{Klevansky:1992} S. P. Klevansky, Rev. Mod. Phys. \textbf{64}, 649 
   (1992).
\bibitem{Hatsuda:1994} T. Hatsuda, T. Kunihiro, Phys. Rep. \textbf{247}, 221 
   (1994).
\bibitem{Bernard:1993} V. Bernard, A. H. Blin, B. Hiller, U.-G. Mei\ss ner, 
   M. C. Ruivo, Phys. Lett. B \textbf{305}, 163 (1993).
\bibitem{Dmitrasinovic:1990} V. Dmitrasinovic, Nucl. Phys. A \textbf{686}, 
   379 (2001).
\bibitem{Naito:2003} K. Naito, M. Oka, M. Takizawa, T. Umekawa, Progr. Theor. 
   Phys. \textbf{109}, 969 (2003).
\bibitem{Osipov:2006b} A. A. Osipov, B. Hiller, J. da Providencia, Phys. Lett. B\ \textbf{634} (2006), 48. 
\bibitem{Osipov:2007a} A. A. Osipov, B. Hiller, A. H. Blin, J. da Providencia, Annals of Phys. \ \textbf{322} (2007), 2021.
\bibitem{Georgi:1984} A. Manohar, H. Georgi, Nucl. Phys. B \textbf{234}, 189 
   (1984).
\bibitem{Osipov:2001} A. A. Osipov, B. Hiller, Phys. Lett. B \textbf{515}, 458 
   (2001); A. A. Osipov, B. Hiller, Phys. Rev. D \textbf{63}, 
   094009 (2001); A. A. Osipov, B. Hiller, Phys. Rev. D \textbf{64}, 
   087701 (2001).
\bibitem{Osipov:2004b} A. A. Osipov, H. Hansen, B. Hiller, Nucl. Phys. A 
   \textbf{745}, 81 (2004).
\bibitem{Osipov:1985} M. K. Volkov, A. A. Osipov, Sov. J. Nucl. Phys. 
  \textbf{41}, 500 (1985).
\bibitem{Manohar:1986} D. B. Kaplan, A. V. Manohar, Phys. Rev. Lett. 
   \textbf{56}, 2004 (1986).
\bibitem{PDG} K. Nakamura et al. (Particle Data Group), J. Phys. G \textbf{37}, 
   075021 (2010). 
\bibitem{Volkov:1984} M. K. Volkov, Ann. Phys. \textbf{157}, 282 (1984).
\bibitem{Volkov:1986} M. K. Volkov, Fiz. Elem. Chastits At. Yadra \textbf{17}, 
   433 (1986).
\bibitem{Su:2007} M. X. Su, L. Y. Xiao, H. Q. Zheng, Nucl. Phys. A 
   \textbf{792}, 288 (2007).
\bibitem{Parganlija:2013} 
 D. Parganlija, P. Kovacs, G. Wolf, F. Giacosa, D. H. Rischke, Phys. Rev. D {\bf 87}, 014011 (2013). 
 \bibitem{Beveren:1986} E. van Beveren, T. A. Rijken, K. Metzger, C. Dullemond,
   G. Rupp, J. E. Ribeiro, Zeit. Phys. C \textbf{30}, 615 (1986).
\bibitem{Jaffe:1977} R. J. Jaffe, Phys. Rev. D \textbf{15}, 267 (1977).
\bibitem{Schechter:2008} A. H. Fariborz, R. Jora, J. Schechter, Phys. Rev. D
   \textbf{77}, 094004 (2008).
\bibitem{Schechter:2009} A. H. Fariborz, R. Jora, J. Schechter, Phys. Rev. D 
   \textbf{79}, 074014 (2009). 
\bibitem{Close:2002} F. E. Close, N. A.T\"ornqvist, J. Phys. G: Nucl. Part. 
   Phys. \textbf{28}, R249 (2002);
\end{thebibliography}
\end{document}